\documentstyle[12pt]{article}
\title{Statistical Entropy of Magnetic Black Holes from Near-Horizon
Geometry}
\author {Mikhail Z. Iofa\thanks{E-mail address: iofa@theory.npi.msu.su}
and Leopoldo A. Pando Zayas\thanks{E-mail address: leopoldo@grg1.phys.msu.su}\\
Nuclear Physics Institute \\
Moscow State University\\
Moscow 119899, Russia }
\date{}

\def\o{\omega}
\def\O{\Omega}
\def\d{\delta}
\def\e{\epsilon}
\def\t{\tau}
\def\p{\phi}
\def\S{\Sigma}
\def\r{\rho}
\def\hhr{\hat{r}}
\def\hhp{\hat{\phi}}
\def\hht{\hat{t}}

\begin{document}

\maketitle

\begin{abstract}
Four-dimensional magnetic black holes including dilaton and abelian gauge
fields which  are solutions of supergravity can also be obtained by
dimensional reduction of the Einstein-Maxwell gravity in five dimensions. In
the extremal case the five-dimensional solutions have horizon and their 
near-horizon geometry is $AdS_3\times S^2 $. In the non-extremal case the 
near-horizon geometry is shown to be the product of the three-dimensional
Ba\~nados-Teitelboim-Zanelli black hole and $S^2$. This allows
to perform  microscopic counting of statistical entropy of magnetic
black holes.  Exact agreement with the geometric entropy is found. 
 The microstates responsible for statistical entropy are located 
in the near-horizon region. 
\end{abstract}


\vfill
\noindent NPI-MSU-98-38/489\\

\section{Introduction} 

Recently considerable attention has been paid to the
problem of the microscopic calculation of the entropy of black holes 
by reducing
the problem to counting the number of microstates 
of the Ba\~nados-Teitelboim-Zanelli (BTZ) 
black hole \cite{btz}. The
counting \cite{carlip} is based on the fact that diffeomorphisms 
which preserve the asymptotic of the $AdS_3$ vacuum solution at spatial
infinity are generated by two copies of the Virasoro algebra with a certain
central charge \cite{hen}. It appears that some non-extremal solutions of 
the type-IIA
supergravity can be reduced via the U-duality transformations to different
forms, including those which under dimensional reduction 
  yield non-extremal 4D (5D) black
holes  and the others which reduce to the product of 
the non-extremal BTZ black hole
and  2 (3)-dimensional sphere parts \cite{udual}. Using these
transformations it is 
possible to connect the geometric entropy to statistical entropy 
calculated by counting the number of states in a conformal
theory associated with the BTZ black hole \cite{sfetsos}. Characteristic 
features of these
solutions are: (i) they are exact; (ii) the BTZ and the sphere parts
completely decouple and are exact classical solutions. 

A generalization of these  results was proposed in \cite{stro} 
where it was
argued that calculation of statistical entropy  can be carried out
equally well for any black hole
whose near-horizon geometry is locally $AdS_3$. 
A realization of this idea
was presented in \cite{bss,bir}, where it was shown that there exist
compactifications of heterotic string theory which yield supersymmetric
solutions  containing $AdS_3$ as a factor. An explicit example
of such construction was presented also in \cite{lar} where, starting
from a configuration of  5M-branes, the geometric entropy of 
4D black holes
was related to the entropy of the BTZ black hole by noting that by 
dimensional reduction the near-horizon region of the solution 
transforms to the
product  $BTZ\times S^2$. In this case, in contrast to the solutions
obtained by means of the U-duality transformations or the heterotic theory
compactifications  \cite{bss}, the $BTZ\times S^2$ form of 
the solution is only approximate in a sense that it is valid only in the 
near-horizon  region $(Q/r << 1)$.

In the present paper we discuss the well-studied example of 4D 
dilaton gravity coupled
to an abelian  gauge field. The equations of motion admit a magnetically
charged black hole; both extremal and non-extremal solutions are available
\cite{gib}.
In the extremal case,  the metric and dilaton are singular at the horizon.
However, a remarkable property of the solution is that, for certain values
of the dilaton coupling, the singularity is resolved by reinterpreting the
solution  as an object in a higher-dimensional spacetime \cite{gib}.  Moreover, 
it appears that in a  neighborhood of the horizon, the higher-dimensional
metric is of the form $AdS_3 \times S^2$. In the non-extremal case, similar
reinterpretation of the four-dimensional solution in terms of a 
higher-dimensional one shows that in
a neighborhood of the horizon the geometry  is  of the form 
$BTZ\times S^2$. This suggests a possibility to compare the
Bekenstein-Hawking entropy of the (non-extremal) 4D magnetic black hole with
statistical entropy  obtained by counting  states in a  conformal theory 
associated with the BTZ black hole. 

\section{Magnetic black hole solution} 

A class of magnetic black hole solutions is obtained by solving the equations
of motion of the action
\begin{equation}
\label{act}
S_4=\frac{1}{16\pi G_4}\int d^4 x \sqrt{-g}\{ R^{(4)} -2(\partial \phi)^2
-e^{-2a\phi}F_2^2\},
\end{equation}
where $a=\sqrt{\frac{p}{p+2}},\quad p =0,1,\ldots $.

We shall be interested in the following solution
\begin{eqnarray}
\label{bh}
ds_4^2 &=&-(1-\frac{r_+}{r})(1-\frac{r_-}{r})^{1/2}dt^2 +
(1-\frac{r_+}{r})^{-1}(1-\frac{r_-}{r})^{-1/2}dr^2 
+ r^2(1-\frac{r_-}{r})^{1/2}d\Omega_2^2\nonumber \\
e^{a\phi}&=&(1-\frac{r_-}{r})^{-1/4} \nonumber \\
F_2&=& Q\epsilon_2\nonumber \\
Q^2&=&\frac{3}{4}r_-r_+.
\end{eqnarray}
Here  $r_+$ and $r_-$ are the outer and inner horizons, the extremal limit
being $r_+=r_-=\mu$ and $\epsilon_2$ is the volume two-form on the unit
2-sphere, $a=\frac{1}{\sqrt{3}}$.  

For $a=\frac{1}{\sqrt{3}}$ the solution (\ref{bh}) can be also obtained 
by dimensional reduction of a solution of the pure Einstein-Maxwell 
action in five dimensions. 
\begin{equation}
\label{em}
S_5=\frac{1}{16\pi G_5}\int d^5\sqrt{-g^5}\{R^{(5)}-F_2^{(5)2}\}
\end{equation}
This action admits a solution of the form 
\begin{eqnarray}
\label{bs}
ds_5^2&=&-(1-\frac{r_+}{r})dt^2 +(1-\frac{r_-}{r})dy^2+
(1-\frac{r_+}{r})^{-1}(1-\frac{r_-}{r})^{-1}dr^2 +r^2d\Omega_2^2\nonumber\\
F_2^{(5)}&=&Q\epsilon_2,
\end{eqnarray}
where $Q$ and $\epsilon_2$ are the same as in (\ref{bh}). Introducing
a new variable $\o$ according to $1-\frac{r_+}{r}=\o^2$, we rewrite the
metric (\ref{bs}) as 
\begin{equation}
\label{ome}
ds_5^2=-\o^2dt^2 + (\delta+\o^2)\frac{r_-}{r_+}dy^2 +\frac{dw^2}{\delta
+\o^2}\frac{4r_+^3}{r_-}\frac{1}{(1-\o^2)^4}+
\frac{r_+^2}{(1-\o^2)^2}d\O_2^2,
\end{equation}
where $\delta=\frac{r_+}{r_-}-1$. In the extremal limit, $r_+ = r_-$ and
$\delta = 0$. In this case the near-horizon geometry is that of the $AdS_3$
space \cite{gib}. Analytically continuing $\o$ to complex
values, i.e., changing $\o \to i\o$, we transform (\ref{ome}) to 
\begin{equation}
\label{bslast}
ds_5^2=-(\o^2-\d)\frac{r_+}{r_-}dy^2+\o^2dt^2+\frac{d\o^2}{\o^2-\d}
\frac{4r_+^3}{r_-}\frac{1}{(1+\o^2)^4}+\frac{r_+^2}{(1+\o^2)^2}\d\O_2^2.
\end{equation}
It is seen that near the horizon $r \approx r_+$, i.e. at small $\o$, the metric
(\ref{bslast}) describes the product of the BTZ black hole with $S^2$,
provided we interpret $y$ as the non-compact and $t$ as the compact
coordinates. To make the correspondence with the BTZ black hole more
transparent we identify 
\begin{equation}
\label{ide}
\d=M, \quad \o^2=\frac{r^2}{l^2}, \quad dy=(\frac{r_+}{r_-})^{1/2}d\t,
\quad \o^2dt^2=r^2d\phi^2, \quad l^2 = \frac{4r_+^3}{r_-}
\end{equation}
and finally obtain
\begin{equation}
ds_5^2=-\frac{r^2-Ml^2}{l^2}d\t^2+r^2d\p^2+\frac{l^2dr^2}{r^2-Ml^2}
\frac{1}{(1+\frac{r^2}{l^2})^4}+\frac{r_+^2d\O_2^2}{(1+\frac{r^2}{l^2})^2}.
\label{fin}
\end{equation}

\section{Bekenstein-Hawking entropies.} 

Let us calculate and compare the
geometric entropy of the magnetic black hole (\ref{bh}) with that of the BTZ
black hole. Using the metric (\ref{bh}), we obtain for the geometric
entropy the following expression:

\begin{equation}
S^{(4)}=\frac{4\pi r_+^2(\frac{r_+}{r_-})^{1/2}\d^{1/2}}{4G_4}.
\end{equation}  
Here $G_4$ is the four-dimensional Newton constant. The BTZ black hole is
a solution to the equations of motion of the three-dimensional gravity
described by the action 
\begin{equation}
\label{s3}
S_3=\frac{1}{16\pi G_3}\int d^3x \sqrt{-g^{(3)}}(R^{(3)}+\frac{2}{{\hat{l}}^2})
\end{equation}
yielding the entropy
\begin{equation}
S^{(3)}=\frac{2\pi \hat{l}M^{1/2}}{4G_3}.
\end{equation}
For a special class of the five-dimensional metrics, to which belongs the
metric (\ref{bs}), the Einstein-Maxwell action (\ref{em}) reduces to the
action (\ref{act}). The correspondence requires the identification 
\begin{equation}
\frac{1}{G_4}=\frac{1}{G_5}\int dy.
\end{equation}
For the geometries which are approximately the product of a three-dimensional
part and $S^2$, the five-dimensional Ricci scalar $R^{(5)}$ is the sum of
$R^{(3)}+R^{(2)}$, where $R^{(2)}=2/r_+^2$. The term $F_2^2=(Q\e_2)^2$ is
equal to $\frac{3}{2}\frac{r_+r_-}{r_+^4}$. Thus, the action (\ref{em})
reduces to 
\begin{equation}
\label{cosmo}
S_5=\frac{1}{16\pi G_5}4\pi r_+^2\int d^3\sqrt{-g^{(3)}}(R^{(3)}
+\frac{2}{r_+^2}-\frac{3}{2}\frac{r_+r_-}{r_+^4})
\end{equation}
which should be compared with (\ref{s3}) giving 
\begin{equation}
\frac{1}{G_3}=\frac{1}{G_5}4\pi r_+^2.
\end{equation}

Now let us 
consider the near-extremal case for which $\frac{r_+}{r_-}\to 1$ and
$\d<<1$. From the identification (\ref{ide}) we have $l^2\approx 4r_+^2$.
On the other hand, independent of this identification, 
the scale $\hat{l}$ is defined by the cosmological constant which enters the
three-dimensional gravity action.
From (\ref{cosmo})  in the near-extremal case we obtain 
$$\frac{2}{r_+^2}(1-\frac{3r_+r_-}{4 r_+^2}) \approx \frac{2}{{\hat{l}}^2}.$$
and thus ${\hat{l}}^2\approx 4r_+^2$.
The correspondence between expressions for  $l^2$ and  ${\hat{l}}^2$  
further supports the self-consistency of our approach.

Finally, taking $y$ to be a compact variable on the circle of the radius $l$,
we find that the geometric entropies $S^{(3)}$ and  $S^{(4)}$ are equal to
each other \footnote{Although there are no {\it a priori} limitations on the
radius of the circle, it would be desirable to understand the reason of this
choice.}. 

\section{Counting the black-hole microstates.} 

Having established the
correspondence between the three and four-dimensional Bekenstein-Hawking
entropies, now we address to the question of the 
microscopic calculation of the
entropy. As above, we discuss the near-extremal case. Let us consider the
near-horizon region $\sqrt{M}<< \frac{r}{l}<< 1$, where the
configuration is approximately the product of the BTZ black hole and $S^2$.
The condition for validity of the semi-classical description of the black
hole $l >> G_3$ is consistent with the restriction 
to the near-horizon  region
if $\frac{l}{G_3} >> \frac{1}{\sqrt{M}} >> 1$.

A puzzle that arises in understanding the origin of the entropy of the BTZ
black hole is that, although the entropy is physically associated with the
near-horizon degrees of freedom, it can be calculated from the Virasoro
algebra connected with the diffeomorphisms 
that preserve the asymptotic behavior
of the metric at large $r$ \cite{stro}. This could have seriously damaged
our discussion because on the one hand, the initial metric (\ref{ome}) was
defined for $\o^2<1$, and on the other hand, the metric (\ref{fin}) can be
interpreted as the product of the BTZ part and $S^2$ only for small $r$.

However, the process of counting of states can be performed using a 
general result concerning the
Chern-Simons action defined on a manifold with topology $\S_2\times R$
\cite{ban, ban1}. In
the three-dimensional Chern-Simons theory there are no local degrees of
freedom, the only relevant degrees of freedom being global charges which
 generate  residual gauge transformations. For a special, but
general enough to
include most of interesting cases, set of boundary conditions and
assumptions on the form of the parameters of residual 
gauge transformations at
the boundary, the global charges form an algebra. For a particular class of
the boundary conditions this algebra  contains the Virasoro algebra. 

In the three-dimensional gravity with a negative cosmological constant
represented as the Chern-Simons theory for the $AdS_3$ space, there appear two
chiral copies of the Virasoro algebra. The Virasoro algebra
written in the standard form has the classical central charge 
$c=\frac{3l}{2G_3}$. 
The "mass" of the BTZ black hole is related to the zero modes of the Virasoro
generators $L_0$ and $\bar{L}_0$ as $M=\frac{8G_3}{l}(L_0+\bar{L}_0)$. This
provides the microscopic entropy which was found to be equal to the
Bekenstein-Hawking entropy \cite{carlip}  calculated  using the Cardy's 
formula for the asymptotic density of states.

In the chain of transformations of the metric $ds_5^2$ which lead to the
metric (\ref{fin}), we performed the transformation $\o \rightarrow i\o$.
Let us discuss this issue first from the point of view of the  $AdS_3$
theory \cite {carlip}. Starting from the $SL(2,R)$ theory in which the 
group element is parameterized as
\begin{equation}
\label{E1}
g=\left(
\begin{array}{ll}
x_1+x_2&x_3+x_0\\
x_3-x_0&x_1-x_2
\end{array}
\right),
\end{equation}
where
\begin{equation}
\label{E2}
x_0^2 +x_1^2 -x_2^2 -x_3^2 =1,
\end{equation}
one can consider (\ref{E2}) as the embedding equation of  $AdS_3$
in a flat space. The universal covering space of  $AdS_3$ can be covered by
three  patches. Since the discussion is the same for any of them, let
us, for definiteness, consider the region parameterized as 
\begin{eqnarray}
\label{E3}
x_1&=&r \cosh\phi \qquad x_0=\sqrt{r^2-1}\sinh t \nonumber \\
x_2&=&r\sinh\p \qquad x_3=\sqrt{r^2-1}\cosh t,
\end{eqnarray}
where $r^2 > 1,\, -\infty < t,\p < \infty$. In every patch the group metric
is
\begin{equation}
\label{E4}
ds^2=-(r^2-1)dt^2 +r^2 d\p^2 +(r^2-1)^{-1}dr^2.
\end{equation}
Changing $r\to r'= ir$ and introducing a new variable $\r$ as $\r^2=r'^2+1$
 we see  from (\ref{E3}) that, 
supplemented with the transformation 
$\p\leftrightarrow t$ the change leaves the metric (\ref{E4}) invariant. 
Thus, we conclude
that this transformation is a symmetry of the theory. 

The BTZ black hole can be generated from the $AdS_3$ solution by the change of
variables 
\begin{equation}
r^2=\frac{\hhr^2-r_-^2}{r_+^2-r_-^2}, \quad t=r_+\hht-r_-\hhp, \quad
\p=-r_-\hht+r_+\hhp, \quad M=r_+^2 +r_-^2, \quad J=2r_+r_-,
\end{equation}
yielding 
\begin{equation}
ds^2=-(\hhr^2-M)d\hht^2 -Jd\hht d\hhp+\hhr^2d\hhp^2
+(\hhr^2-M+\frac{J^2}{4\hhr^2})^{-1}d\hhr^2
\end{equation}
supplemented with the identification 
\begin{equation}
\hhp\sim \hhp+2\pi.
\end{equation}

Since the $AdS_3$ solution was argued to be invariant under the
transformation $r\to i\sqrt{\r^2-1}$, one can interchange $t$ and $\p$ in
the change of variables leading to the BTZ solution and simultaneously
regard $t$ as a compact variable. 
 
\section{Discussion}

Microscopic calculations of the entropy of the black hole usually start from
 construction of a configuration of the black hole in terms of the exact 
solutions of 10D or 11D supergravities. In this paper our starting point 
was a  simple 4D
dilatonic supergravity interacting with the abelian gauge field which can be
lifted to a  5D Einstein-Maxwell gravity. No construction in terms
of intersecting M-branes was used.

In examples in which the BTZ black hole appears as a factor in 10D (11D)
configuration decoupled from the remaining part of solution obtained by
application of the U-duality transformations there is no natural scale for a
definition of a near-horizon region. Separation of the 
near-horizon region appears if there is no exact decoupling of 
the 3D black hole
 from the remaining part of the configuration; the decoupling parameter being
also the scale which defines the near-horizon region; this refers both to 
the examples of intersecting branes \cite{lar} and to the present case. 
Moreover, it appears that microscopic degrees of freedom responsible 
for the entropy of the black hole are located in the near-horizon region.

The result of papers \cite{ban, ban1}, quoted in the preceding section 
in connection with
the problem of counting of the microstates concerning the appearance of the
Virasoro algebra, do not require consideration of the asymptotic region
$r \rightarrow \infty$ and is valid for an arbitrary boundary surface
located at a finite distance from the horizon.

\begin{center}
{\large \bf Acknowledgments}
\end{center}
This work was  partially supported by the RFFR  grant No 98-02-16769

\pagebreak


\begin{thebibliography}{99}

\bibitem{btz}
M.~Ba\~nados, C.~Teitelboim and J.~Zanelli,   Phys. Rev. Lett.  69 (1992)
1849, hep-th/9204099;\\
M.~Ba\~nados, M.~Henneaux, C.~Teitelboim and J.~Zanelli,
 Phys. Rev. D 48 (1993) 1506, gr-qc/9302012.

\bibitem{carlip}
S.~Carlip, Phys. Rev. D 51 (1995) 632, gr-qc/9409052;\\ 
 Phys. Rev. D 55 (1997) 878, gr-qc/9606043;\\ 
 Nucl. Phys. Proc. Suppl. 57 (1997) 8, 
gr-qc/9702017.

\bibitem{hen}
J. D.~Brown and M.~Henneaux, 
 J. Math. Phys. 27 (1986) 489;\\
 Commun. Math. Phys.104 (1986) 207.

\bibitem{udual}
S.~Hyun, U-duality between Three and Higher Dimensional Black Holes
hep-th/9704005;\\
H. J.~Boonstra, B.~Peeters and K.~Skenderis, 
 Phys. Lett. B 411 (1997) 59, hep-th/9706192; \\ Branes and anti-de
Sitter Spacetimes, hep-th/9801076.

\bibitem{sfetsos}
K.~Sfetsos and K.~Skenderis, Nucl.Phys. B 517 (1998) 179-204, hep-th/9711138.

\bibitem{stro}
A.~Strominger, Black hole entropy from near horizon microstates, hep-th/9712251.

\bibitem{lar}
V.~Balasubramanian and F.~Larsen, Near Horizon Geometry and Black Holes in
Four Dimensions, hep-th/9802198.

\bibitem{gib}
G. W.~Gibbons, G. T.~Horowitz and P. K.~Townsend, 
 Class.Quant.Grav. 12 (1995) 297, hep-th/9410073.

\bibitem{bss}
D.~Birmingham, I.~Sachs, and S.~Sen, Entropy of Three-Dimensional 
Black Holes in String Theory, hep-th/9801019.

\bibitem{bir}
D.~Birmingham, String Theory Formulation of anti-de Sitter Black Holes,
hep-th/9801145.

\bibitem{ban}
M.~Ba\~nados, Phys. Rev. D 52 (1995) 5816, hep-th/9405171.

\bibitem{ban1}
 M.~Ba\~nados, T.~Brotz and M.~Ortiz, Boundary dynamics and the statistical 
mechanics of the 2+1 dimensional black hole, hep-th/9802076. 



\end{thebibliography}
\end{document}